\documentclass{elsart1p}

\usepackage{amssymb}

\usepackage[dvips]{graphicx,color}

\journal{Physics Letters B}

\begin{document}

\begin{frontmatter}

\title{Relativistic Effects in Exclusive pd Breakup Scattering at Intermediate Energies}

\author[Lin]{T.~Lin},
%
\author[Lin,Elster]{Ch.~Elster},
%
\author[Polyzou]{W.~N.~Polyzou},
%
\author[Gloeckle]{W.~Gl\"ockle},
%
\address[Lin]
{Institute of Nuclear and Particle Physics,  and
Department of Physics and Astronomy,  Ohio University, Athens, OH 45701, USA}
\address[Elster]
{Physics Division, Argonne National Laboratory, Argonne, IL 60439, USA}
\address[Polyzou]
{Department of Physics and Astronomy, The University of Iowa, Iowa City,
IA 52242, USA}
\address[Gloeckle]
{Institute for Theoretical Physics II, Ruhr-University Bochum,
D-44780 Bochum, Germany}




\begin{abstract}
The relativistic Faddeev equation for three-nucleon scattering is
formulated in momentum space and directly solved in terms of momentum
vectors without employing a partial wave decomposition. Relativistic
invariance is achieved by constructing a dynamical unitary
representation of the Poincar\'e group on the three-nucleon Hilbert
space.  The exclusive breakup reaction at 508~MeV is calculated based
on a Malfliet-Tjon type of two-body interaction and the
cross sections are compared to measured cross sections at this
energy. We find that the magnitude of the relativistic effects can be
quite large and depends on the configurations considered. In spite of
the simple nature of the model interaction, the experimental cross
sections are in surprisingly good agreement with the predictions of the
relativistic calculations.  We also find that although for specific
configurations the multiple scattering series converges rapidly, this
is in general not the case.

\end{abstract}



\begin{keyword}
Relativistic Quantum Mechanics \sep Faddeev Equation  
\sep The Quantum Mechanical Three-Body Problem \sep n-d Scattering
\PACS 21.45+v \sep 24.10.Jv \sep 25-10.+s
\end{keyword}


\end{frontmatter}


Breakup reactions in the proton-deuteron ($pd$)  system at intermediate
energies have been studied experimentally quite intensively in recent
decades. A prominent set of data can be found in the comprehensive
overview of the experiments completed at Saturne-2~\cite{Saturn}.
However, the theoretical interpretation faced and still faces serious
challenges. At those energies pion production channels are open and
nuclear resonances play a role. In contrast to the energy regime below
the pion threshold, where high precision nucleon-nucleon (NN) forces are
established~\cite{AV18,CDBONN,NIJM}, and nuclear forces based on
effective chiral dynamics are being developed~\cite{Epelbaum},  the
intermediate energy regime, in which the lowest nucleon resonances
play a role, does not yet have forces describing data with comparable
quality~\cite{Pricking:2007qx}.  Moreover it can be expected that
three-nucleon (3N) forces will play a more prominent role 
in the intermediate energy regime. 
The tendency for an increasing
importance of 3N forces with increasing energy has already been seen in  3N
observables at low energies~\cite{Witala01}.

A consistent treatment of intermediate energy reactions requires a
Poincar\'e symmetric quantum theory \cite{Wigner39}.  In addition, the
standard partial wave decomposition, successfully applied below the
pion-production threshold~\cite{wgphysrep}, is no longer an adequate
numerical scheme due to the proliferation of the number of partial
waves.  Thus, the intermediate energy regime is a new territory for
few-body calculations, which waits to be explored.

The aim of this article is to address two aspects in that list of challenges:
exact Poincar\'e invariance and calculations using vector
variables instead of partial waves.  
In a series of publications~\cite{Hang} the technique
of solving the nonrelativistic momentum-space Faddeev equation without
partial waves has been mastered, for bound as well as scattering
states.  The Faddeev equation, based on a Poincar{\'e} invariant mass
operator, has been formulated in detail in~\cite{Lin:2007ck}.  The
resulting Faddeev equation has both kinematical and dynamical
differences with respect to the corresponding nonrelativistic equation.

The formulation of the theory is given in a representation of
Poincar\'e invariant quantum mechanics where the interactions are
invariant with respect to kinematic translations and rotations
\cite{Coester65}.  The model Hilbert space is a three-nucleon Hilbert
space (thus not allowing for absorptive processes).  The method used
to introduce the NN interactions in the unitary
representation of the Poincar\'e group allows to input
high-precision NN interactions~\cite{AV18,CDBONN,NIJM} in a way
that reproduces the measured two-body observables. However in this
study we use a simpler interaction consisting of a superposition of 
an attractive and a
repulsive Yukawa interaction that supports a bound state with the
deuteron binding energy.  Poincar\'e invariance and $S$-matrix cluster
properties dictate how the two-body interactions must be embedded in
the three-body dynamical generators.  Scattering observables are
calculated using Faddeev equations formulated with the mass Casimir
operator (rest Hamiltonian) constructed from these generators.

In \cite{Lin:2007ck} the driving term in the relativistic Faddeev
equation (first order in the two-body transition operator) has been
used to evaluate $pd$ elastic scattering as well as break-up cross
sections.  This has now been completed by fully solving the
relativistic Faddeev equation based on the numerical techniques previously
used to solve the nonrelativistic Faddeev equation~\cite{Hang}. Our
calculations converge well up to 2 GeV, indicating the applicability
of the formulation of the Faddeev equation based on vector variables for 
intermediate energies.
We want to point out that the relativistic Faddeev equations with 
realistic spin-dependent
interactions have been solved below the pion-production threshold
in \cite{Witala} a partial wave basis.  

In order to estimate the size of relativistic effects the interactions
employed in the nonrelativistic and relativistic calculations presented
here are chosen to be phase shift equivalent.  This is achieved in this article
by adding the interaction to the square of the mass 
operator~\cite{cps,Keister:2005eq}.  In this comparison differences in
the relativistic and nonrelativistic calculations first appear in the
three-body calculations.  Those differences are in the choice of kinematic
variables (Jacobi momenta are constructed using Lorentz boosts rather
than Galilean boosts) and in the embedding of the two-body
interactions in the three-body problem, which is a consequence of the
non-linear relation between the two and three-body mass operators.
These differences modify the permutation operators and the off-shell
properties of the kernel of the Faddeev equations.

The relativistic Faddeev equation is solved in the form
\begin{equation}
T(z) = t(z) P + t(z) P( z-M_0)^{-1} T(z),
\label{eq:1}
\end{equation}
where $P$ is a permutation operator and $t(z)$ is
the two-body transition operator embedded in the three-nucleon
Hilbert space defined as the solution to
\begin{equation}
t(z) = V + V ( z-M_0)^{-1} t(z).
\label{eq:2}
\end{equation}
Here the interaction is given as $ V= M-M_0$, where $M$ 
is the three-particle mass operator with one
two-body interaction alone and $M_0$ the corresponding free mass operator.  The
physical transition amplitudes for elastic scattering and the breakup
processes can be expressed in terms of $T(z)$.  
Our aim in this
article is to calculate exclusive three-body breakup
processes in the intermediate energy region and compare
relativistic and nonrelativistic results. Since we generate the
solution of the Faddeev equation from the multiple scattering series
(resulting from iterating Eq.~(\ref{eq:1})), we can also obtain insight into the
contributions of the various orders of that series.  The exact
solution to the Faddeev equation is constructed from the multiple
scattering series using Pad\'e summation.  

Though our two-body force is simple, we want to compare to a $^2$H(p,2p)n
experiment at a projectile kinetic energy 508~MeV~\cite{Punjabi:1988hn}
to see if our calculation captures essential features of the measurement.
Differences in the predictions of our relativistic and nonrelativistic 
calculations are already very pronounced at this energy.

The five-fold laboratory differential cross section for exclusive scattering
is given by \cite{breakupcorr}
\begin{eqnarray}
\frac {d^5\sigma_{br}^{lab}}{ d\Omega _1d\Omega _2dE_1} 
&=& (2\pi)^4 \frac{E(q_0) E_d(q_0)  E(q)}{2k_{lab}m_d } 
\frac{p_1p_2^2} {p_2({\sf E}-E(p_1)) - E(p_2)
(\mathbf{P}-\mathbf{p}_1) \cdot \hat {\mathbf{p}}_2 } 
\nonumber \\
& & \times E(k) \sqrt{4 E^2(k)+\mathbf{q}^2} \;
\left| \langle \mathbf{k}, \mathbf{q}  \Vert U_0 \Vert
\varphi_d, \mathbf{q}_0 \rangle  \right| ^2 .
\label{eq:3}
\end{eqnarray}
Here $U_0=(1+P)T$ is the operator for breakup scattering, ${\sf E}$ is the
total energy of the system and $\mathbf{P}$ its total momentum.

\begin{figure}
\begin{center}
\includegraphics[width=11cm]{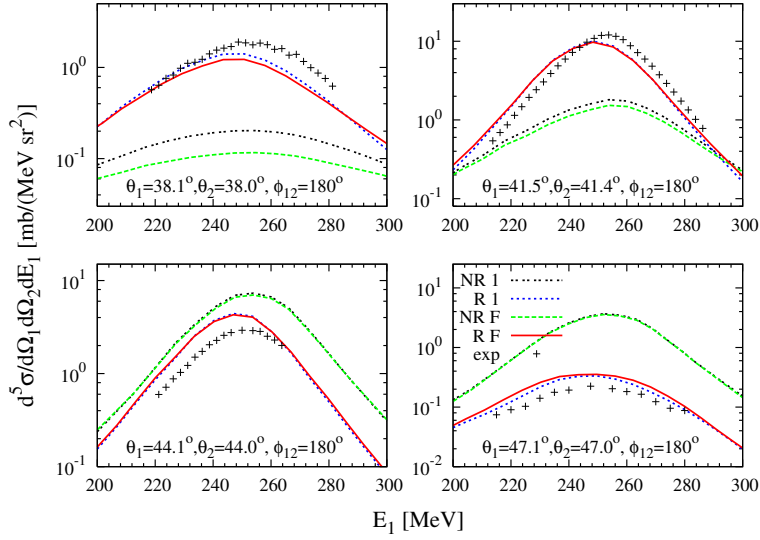}
\end{center}
\caption{The exclusive differential cross section for the $^2$H(p,2p)n
reaction at 508~MeV laboratory projectile energy  for different 
proton angle pairs $\theta_1$-$\theta_2$ symmetric around the beam axis
as function of the laboratory kinetic energy of 
one of the outgoing protons. The solid line (R~F) represents the full
relativistic solution of the Faddeev equation, while the dotted curve (R~1)
indicates the relativistic calculation based on the 1st order in the multiple
scattering expansion of the Faddeev amplitude. The corresponding nonrelativistic
full solution of the Faddeev equation is given by the short-dashed curve (NR~F)
and its 1st order contribution by the double-dotted curve (NR~1). The data are
taken from Ref.~\protect\cite{Punjabi:1988hn}. 
\label{fig1}}
\end{figure}

\begin{figure}
\begin{center}
\includegraphics[width=11cm]{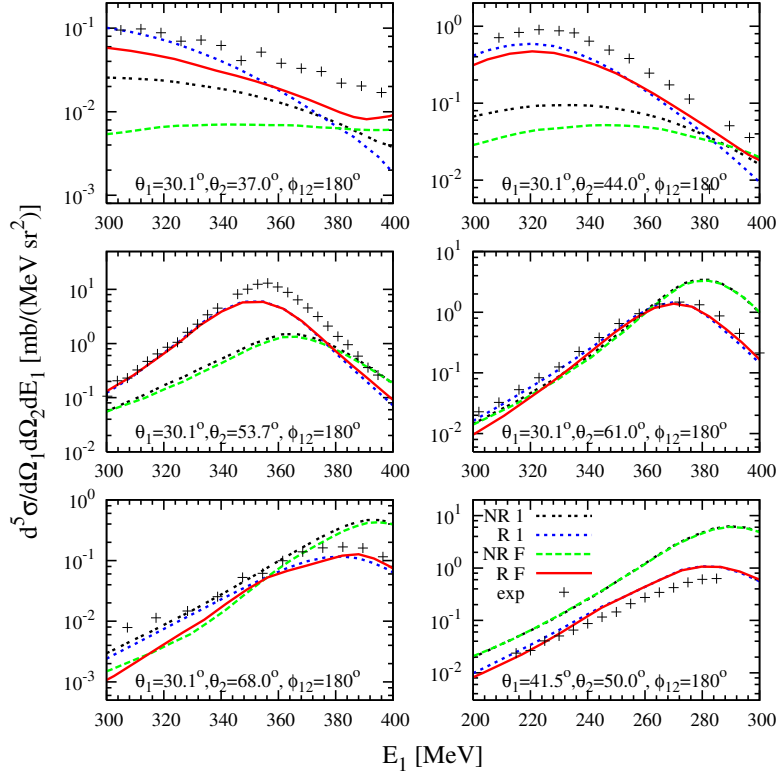}
\end{center}
\caption{The exclusive differential cross section for the $^2$H(p,2p)n
reaction at 508~MeV laboratory projectile energy  for different 
proton angle pairs $\theta_1$-$\theta_2$ asymmetric around the beam
axis as function of the laboratory 
kinetic energy of one of the outgoing protons.
The meaning of the curves is the same as in Fig.~\ref{fig1}. The data are taken
from Ref.~\protect\cite{Punjabi:1988hn}.
\label{fig2}}
\end{figure}

\begin{figure}
\begin{center}
\includegraphics[width=11cm]{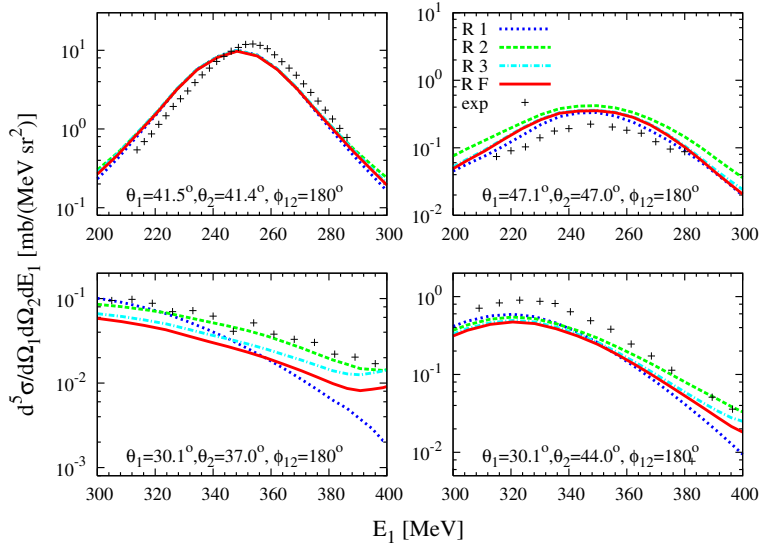}
\end{center}
\caption{The exclusive differential cross section for the $^2$H(p,2p)n
reaction at 508~MeV laboratory projectile energy  for different 
proton angle pairs $\theta_1$-$\theta_2$ as function of the laboratory 
kinetic energy of one
of the outgoing protons. The solid line (R~F) represents the full
relativistic solution of the Faddeev equation. The dotted line (R~1) represents
the relativistic calculation based on the 1st order term in the multiple
scattering expansion of the Faddeev amplitude, while the short-dashed curve (R~2) 
includes the first two terms, and the dash-dotted curve (R~3) the first three
terms.  The data are taken from Ref.~\protect\cite{Punjabi:1988hn}.
\label{fig3}}
\end{figure}

\begin{figure}
\begin{center}
\includegraphics[width=11cm]{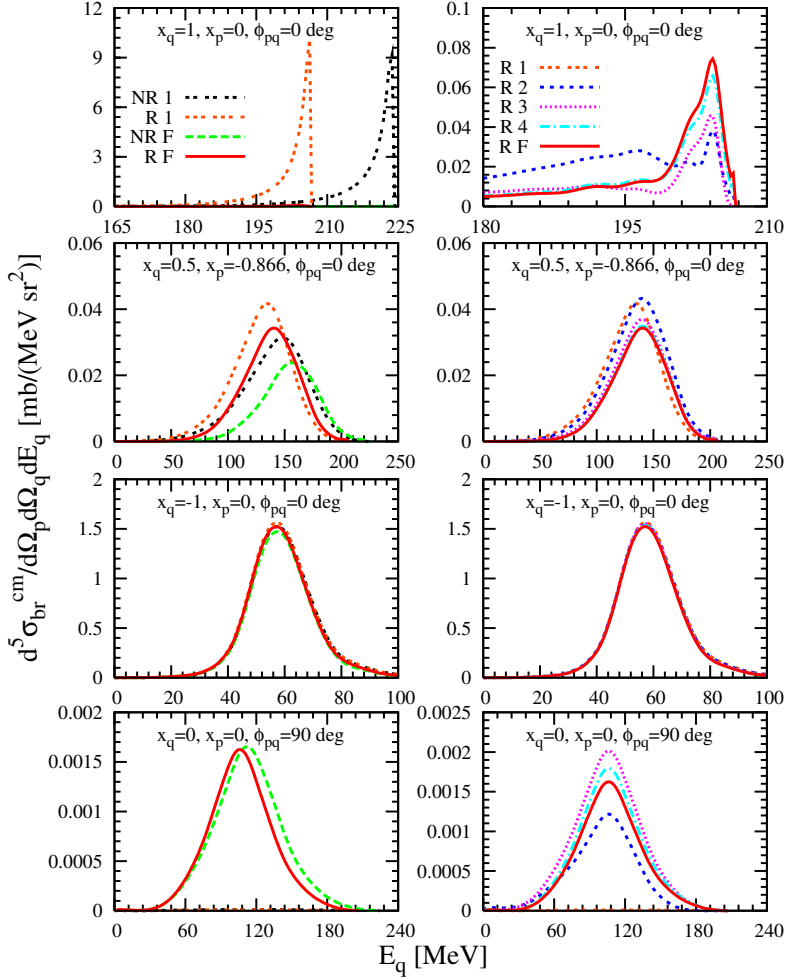}
\end{center}
\caption{The exclusive differential cross section for the $^2$H(p,2p)n
reaction at 508~MeV laboratory projectile energy for different star
configurations as function of the c.m. kinetic energy of one of the outgoing
protons for different c.m. angle pairs $x_q =\cos \theta_q$, $x_p = \cos
\theta_p$. The angle $\phi_{pq}$= 0~deg indicates coplanar star configurations, while 
$\phi_{pq}$ = 90~deg indicates a space star.
All figures show the full relativistic Faddeev calculation as solid line (R~F). 
The left column compares the full relativistic calculation with the full
nonrelativistic one (NR~F), as well as the corresponding relativistic (R~1) and
nonrelativistic (NR~1) first order calculations. The right column shows the
subsequent sum of the lowest orders in the multiple scattering series contributing 
to the full relativistic result. The meaning of those curves is the same as in
Fig.~\ref{fig3}.
\label{fig4}} 
\end{figure}

The results of our calculations are displayed in Figs.~\ref{fig1}-\ref{fig4}.
In Fig.~\ref{fig1} 
we chose selected outgoing proton laboratory angle pairs $\theta_1-\theta_2$ 
from Ref.~\cite{Punjabi:1988hn}, which are
symmetric around the beam axis.  The cross
sections are plotted against the laboratory kinetic energy of one of the outgoing
protons. It is interesting to see that for the smaller angle pairs,
$\theta_{1,2} = 38.1^o$ and $41.5^o$, the relativistic cross sections (solid lines)
are considerably larger than the nonrelativistic ones (dashed
lines). For angle pairs
around $\theta_{1,2} = 44^o$ and larger this reverses and the relativistic
cross sections fall below the nonrelativistic ones.  A similar phenomenon has
been observed in Ref.~\cite{Skibinski:2006cg} for a projectile energy
of  200~MeV, where the
relativistic Faddeev equation has been solved based on the CD-Bonn 
NN force. Surprisingly, our present relativistic calculations come close to
the data~\cite{Punjabi:1988hn}. The fact that the first order results nearly
coincide with the full calculation may be explained by the quasi free scattering (QFS)
condition, which however, is realized only for the angle pair
$\theta_{1,2}=41.5^o-41.4^o$. 

Configurations in which the outgoing protons are measured asymmetric with
respect to the beam axis are shown in Fig.~\ref{fig2}.
Again, the relativistic effects are very pronounced in all configurations shown. 
The angle 
combination $30.1^o-53.7^o$ is the only QFS condition.
In order to illustrate how the cross section is built up by the lowest order
terms of the multiple scattering series, we show in 
Fig.~\ref{fig3} the relativistic results for the first order in the two-body
t-matrix, then add successively one and two rescattering terms,
and compare to the full calculation.
For the QFS condition, $41.5^o-41.4^o$, rescattering terms do not play a role, 
whereas in the two asymmetric configurations even the 3rd order in the 
multiple scattering series is not quite sufficient to arrive at the full result
over the entire measured energy range.

To explore the importance of higher order terms in the multiple scattering
series, we display in Fig.~\ref{fig4} 
three configurations for the coplanar star and one for the space star.
(In the star configurations the interparticle pair
angles are 120$^o$ in the center of momentum (c.m.) system).  
The angles $x_q = \hat q \cdot \hat q_0$ and  $x_p = \hat
p\cdot \hat q_0$ refer to the c.m. system with $\bf p$
and $\bf q$ being Poincare-Jacobi momenta and $\bf q_0$ being the initial
particle momentum. The angle $\phi_{pq} = 0^o$ indicates coplanar star
configurations, while $\phi_{pq }= 90^o$ indicates a space star
configuration.  For $x_q = 1$ (one particle being ejected forward along the
direction of ${\bf q_0}$) the first order result is completely misleading 
and about two
orders of magnitude higher than the fully converged result, which in turn
requires more than four orders in the multiple scattering series. 
However, the first order clearly shows the shift in the peak position characteristic
for relativistic calculations~\cite{Lin:2007ck}.
The configuration $x_q = 0.5$ also exhibits this shift in the peak position
when comparing relativistic and nonrelativistic calculation, and again
the full result is built up by several rescattering contributions.
The coplanar star at $x_q = -1$ exhibits the QFS condition (in the laboratory
frame particle $1$ would stay at rest), and thus the first order is
sufficient. In addition there is no shift in the peak position due to
relativistic effects, which can be understood as consequence of the particle 
being at rest. 
In case of the space star, the first order does not contribute,
and the cross section of this configuration is slowly built up by rescattering
contributions, requiring orders higher than four. The star configuration shown is
representative for all space star configurations achieved by rotation in the
plane perpendicular to the beam.

These observations suggest that even at intermediate energies the full
solution of the Faddeev equation is needed to make reliable
predictions.  There is no loss of generality in using this formalism:
because the two-nucleon interactions are fit to two-nucleon data and
$S$-matrix cluster properties are used to embed the two-nucleon
interactions in the three-nucleon system, any other Poincar\'e 
invariant three-nucleon
model with these properties differs from this one by at most a
three-body interaction.  
The difference between the relativistic and nonrelativistic calculations using
the same two-body input first appears in the three-body system.  Our
calculation indicates measurable differences in the corresponding
relativistic and nonrelativistic breakup cross sections.  Even though
our model interaction is much simpler than a realistic two-body
interaction, the exclusive breakup cross section  are in
surprisingly good agreement with existing data at 508 MeV.

\noindent
The results reported in this letter suggest that \\
(a) A direct integration without employing a partial wave decomposition
 can successfully be
  used to solve the relativistic Faddeev equations with embedded NN
  interactions. \\
(b) The formulation of Poincar\'e invariant quantum theory with
  a three-dimen-sional kinematic Euclidean symmetry is a suitable
  formalism for treating scattering in the intermediate energy region.
  This is supported by the surprising agreement with the measurements
  reported in \cite{Punjabi:1988hn}. \\
(c) Convergence of the multiple scattering series is not
  guaranteed at intermediate energies.

\noindent
A detailed description of our results comprising elastic,
inclusive and exclusive processes will follow in a subsequent 
publication.

   


\section*{Acknowledgments}
\vspace{-2mm}
This work was performed in part under the
auspices of the U.~S.  Department of Energy, Office of Nuclear Physics,
under contract No. DE-FG02-93ER40756 with Ohio University,
contract No. DE-FG02-86ER40286 with the University of Iowa, and contract No.
DE-AC02-06CH11357 with Argonne National Laboratory.
The authors would like to acknowledge fruitful discussions with D.R. Phillips
contributing to this work, and thank V. Punjabi for providing  the experimental
data. C.E. acknowledges the hospitality of the Institute of Modern Physics
(IMP) in Lanzhou, China, where this work was initiated.
 We thank the Ohio  Supercomputer Center (OSC) for the use of
their facilities under grant PHS206.



\clearpage
\noindent

\end{document}